\documentclass[twocolumn,astrosymb,twocolappendix]{aastex631}

\usepackage{tablefootnote}

\usepackage[T1]{fontenc}
\usepackage{amsmath}
\usepackage{epigraph}
\usepackage{threeparttable}
\usepackage{soul}
\usepackage{hyperref}

\newcommand{\hide}[1]{}
\newcommand{\angstrom}{\textup{\AA}}

\def\f17{f_{17}}

\def\ergcm2s{\ifmmode {\rm\,erg\,cm^{-2}\,s^{-1}}\else
                ${\rm\,ergs\,cm^{-2}\,s^{-1}}$\fi}
\def\ergsec{\ifmmode {\rm\,erg\,s^{-1}}\else
                ${\rm\,ergs\,s^{-1}}$\fi}
\def\mstar{\ifmmode {M^*_{UV}}\else
                ${M^*_{UV}}$\fi}
\def\phistar{\ifmmode {\phi^*}\else
                ${\phi^*}$\fi}
\def\zo{\ifmmode {12 + \log[{\rm O}/{\rm H}]}\else  
                ${12 + \log[{\rm O}/{\rm H}]}$\fi}
                
\def\lya{\ifmmode {\hbox{Ly}\alpha}\else  
                Ly$\alpha$\fi}

\def\OIII{[\mbox{O\,{\sc iii}}]}

\def\OII{[\mbox{O\,{\sc ii}}]}

\def\Lya{Ly$\alpha$}
\newcommand{\reff}{$r_{\rm eff}$}
\newcommand{\ewlya}{EW(Ly$\alpha$)}

\shorttitle{Distinctly Small Sizes of Ly$\alpha$-emitters}
\shortauthors{Kim et al.}

\begin{document}

\title{Compact Size, High $\Sigma$SFR: Defining Morphological Features of Ly$\alpha$-Emitters}

\correspondingauthor{Keunho J. Kim}
\author[0000-0001-6505-0293]{Keunho J. Kim}
\affil{IPAC, California Institute of Technology, 1200 E. California Boulevard, Pasadena, CA 91125, USA}
\email{keunho11@ipac.caltech.edu}

\author[0000-0002-8630-6435]{Anahita Alavi}
\affiliation{IPAC, California Institute of Technology, 1200 E. California Boulevard, Pasadena, CA 91125, USA}

\author[0000-0002-9593-0053]{Christopher Snapp-Kolas}
\affiliation{Department of Physics $\&$ Astronomy, University of California, Riverside, CA 92521, USA}

\author[0000-0002-4935-9511]{Brian Siana}
\affiliation{Department of Physics $\&$ Astronomy, University of California, Riverside, CA 92521, USA}

\author[0000-0001-5492-1049]{Johan Richard}
\affiliation{Univ Lyon, Univ Lyon1, Ens de Lyon, CNRS, Centre de Recherche Astrophysique de Lyon UMR5574, F-69230, Saint-Genis-Laval, France}

\author[0000-0002-7064-5424]{Harry Teplitz}
\affiliation{IPAC, California Institute of Technology, 1200 E. California Boulevard, Pasadena, CA 91125, USA}

\author[0000-0001-6482-3020]{James Colbert}
\affiliation{IPAC, California Institute of Technology, 1200 E. California Boulevard, Pasadena, CA 91125, USA}

\author[0000-0001-7166-6035]{Vihang Mehta}
\affiliation{IPAC, California Institute of Technology, 1200 E. California Boulevard, Pasadena, CA 91125, USA}

\author[0000-0002-0943-0694]{Ana Paulino-Afonso}
\affiliation{Instituto de Astrofísica e Ciências do Espaço, Universidade do Porto—CAUP, Rua das Estrelas, PT4150-762, Porto, Portugal}

\author[0000-0002-5057-135X]{Eros Vanzella}
\affiliation{INAF – OAS, Osservatorio di Astrofisica e Scienza dello Spazio di Bologna, via Gobetti 93/3, I-40129 Bologna, Italy}



\begin{abstract}
The mechanisms of Ly$\alpha$ photon escape are key to understanding galaxy evolution and cosmic reionization, yet remain poorly understood.
We investigate the UV-continuum sizes of 23 Ly$\alpha$ emitters (LAEs) at Cosmic Noon ($1.7 < z < 3.3$), extending previous size analyses to include fainter galaxies ($M_{\rm UV} \simeq -14$) using gravitational lensing. 
Our results show that these LAEs are unusually small for their luminosity, with a mean effective radius ($r_{\rm eff}$) of $170 \pm 140$ pc.
They follow a distinct size-luminosity relation, with an intercept at $M_{\rm UV} = -21$ approximately three times smaller than typical star-forming galaxies (SFGs) at similar redshifts. 
This relation, however, is consistent with that of low-redshift Green Pea galaxies, suggesting that LAEs maintain compact sizes across redshifts. 
We also find that Ly$\alpha$ equivalent width (EW(Ly$\alpha$)) increases with decreasing $r_{\rm eff}$, confirming previous findings. 
The small sizes of LAEs lead to high star formation surface densities ($\Sigma$SFR $= 1-600 M_{\sun} \ \rm{yr}^{-1} \ \rm{kpc^{-2}}$), clearly separating them from typical SFGs in the $\Sigma$SFR vs. $r_{\rm eff}$ space. 
Given that high $\Sigma$SFR is linked to strong galactic outflows, our findings imply that compact morphology plays a key role in Ly$\alpha$ escape, likely facilitated by outflows that clear under-dense channels in the ISM.
Thus, these results demonstrate that compact size and high $\Sigma$SFR can help identify Ly$\alpha$-emitters.
\end{abstract}



\section{Introduction} \label{sec:intro}
\lya -emitters (LAEs) are a class of galaxies that emit prominent \lya\ photons.  
Since \lya\ emission line is usually generated by intense star formation activity and/or AGN, most LAEs have high star formation rate (SFR) for their stellar mass (i.e., specific star formation rate (sSFR) $\gtrsim 10^{-8} \rm{yr^{-1}}$) and relatively young, light-weighted stellar population ages ($\lesssim 50 \ \rm{Myr}$), as derived from SED modelling. \citep[e.g.,][]{malh02,gawi07,pirz07,fink15a,liu23}.
However, because \lya\ photons are resonantly scattered by neutral hydrogen and can be significantly absorbed by dust \citep[e.g.,][]{ahn03,chan23}, not all star-forming galaxies (SFGs) show \lya\ emission.
This means that for a galaxy to become a LAE, \lya\ photons have to find a way to escape from star-forming regions within the galaxy all the way to the intergalactic medium (IGM).
So, which properties of LAEs make \lya\ escape possible?

Understanding the escape process of \lya\ photons is also important for deciphering the cosmic reionization processes in the early Universe ($z > 6$).
This is because both \lya\ photons and ionizing photons (Lyman-continuum, LyC) generally require low column density of neutral hydrogen ($N(HI)$) and/or low amount of dust along their escape pathways, as clearly supported by radiative transfer models \citep[e.g.,][]{neuf91,verh15,chan23}.
Numerous observations of LyC leakers across redshifts ($0 \lesssim z \lesssim 4$) have confirmed such close connections between \lya\ and LyC emission by finding that the majority of LyC leakers are indeed strong LAEs \citep{verh15,deba16,izot16,izot18a,stei18,izot21,gaza20,pahl21,flur22}.
Thus, studying the physical properties of LAEs and understanding their \lya\ (and potentially LyC) escape mechanisms have been important topics in the fields of galaxy evolution and cosmology, respectively \citep[e.g.,][and references therein]{malh02,malh12,rhoa00,rhoa14,fink15a,oyar17,rive17}.

In particular, at high redshifts $2 \lesssim z \lesssim 7$,  morphology analyses of bright ($M_{\rm UV}\lesssim -18$) LAEs show that the galaxies are mostly compact (with small effective radius $r_{\rm eff} \lesssim 1.5$ kpc), often with clumpy features shown in UV images \citep[e.g.,][]{dowh07,over08,bond09,tani09,gron11,bond12,malh12,jian13,paul18,shib19,rito19,redd22,liu23}.
Based on the approximately constant physical sizes of LAEs over a wide span of redshift ($2 \lesssim z \lesssim 6$), \citet{malh12} suggested that the compact morphology of LAEs is a crucial condition for a galaxy to become a LAE.
This idea is qualitatively consistent and further supported by later analytic calculations on morphologically compact conditions (i.e., small \reff\ and high star formation surface density $\Sigma$SFR) for LAEs and LyC leakers by \cite{cen20}.

Later studies \citep[i.e.,][]{izot16,izot18a,kim20,kim21} extended the morphology analysis of LAEs to low redshift ($z \sim 0.3$) based on a sample of ``Green Pea'' galaxies (GPs). These are a class of local starburst galaxies characterized by a strong \OIII $\lambda 5007$ emission line (which gives them a greenish optical color at their redshift) and unresolved, compact morphology seen in SDSS images \citep{card09}.
The studies confirmed that the typical size of low-$z$ LAEs is similar to those of high-$z$ counterparts rather than showing a redshift-dependent size growth seen in typical (continuum-selected) SFGs \citep[e.g.,][]{shen03,vand14,shib15}, corroborating the idea about the non-evolving characteristic sizes of LAEs and the importance of compact size in \lya\ emission from a galaxy.

In this paper, we extend previous size analyses of LAEs to fainter UV luminosity ($M_{\rm UV} \simeq -14$) by leveraging the gravitational lensing effects of the foreground clusters.
By extending a UV-luminosity range towards faint galaxies, our analysis reveals the morphological properties of faint LAEs and investigates whether the faint LAEs also show distinctly small sizes compared to typical SFGs, as shown in the relative bright range ($M_{\rm UV}\lesssim -18$), or show substantially different morphological properties such as diffuse extended structures.

%We leverage the combined effect of deep HST imaging with strong gravitational effects by foreground galaxy clusters to securely measure the UV-continuum size of our sample of LAEs (See Section \ref{sec:sample and analysis}) down to faint UV luminosity. Our sample of 24 LAEs are at $1.7 < z < 3.3$ and previously identified as \lya -emitters with the Keck spectroscopic observations and presented in \cite{snap23}. 

We also investigate the connection between \lya\ emission properties (e.g., \lya\ equivalent width) and the UV size to study more general aspects of the relation between \lya\ escape and UV continuum size.

Section \ref{sec:sample and analysis} describes our galaxy sample and procedures for UV size and luminosity measurements. 
In Section \ref{sec3:retuls}, we present our results. 
We discuss the size and luminosity properties of our sample LAEs and compare them with other types of galaxies in Section \ref{sec4:Discussion}.
We summarize our conclusions in Section \ref{sec5:summary and conclusions}. 
Throughout this paper, we adopt the AB magnitude system and the $\Lambda$CDM cosmology of ($H_{0}$, $\Omega_{m}$, $\Omega_{\Lambda}$) = (70 $\rm{kms^{-1}}$ $\rm{Mpc^{-1}}$, 0.3, 0.7).

\section{Sample Selection and Analysis} \label{sec:sample and analysis}
\subsection{LAE Sample}
\label{sec 2.1: LAE sample}

%;------2.1 LAE Sample------
We define a sample of 23 LAEs drawn from \cite{snap23}, where the LAEs were identified during the Keck/LRIS spectroscopy follow-up campaign.
This campaign was designed to target photometrically-selected $1.5 < z_{\rm phot} < 3.5$ lensed galaxies with apparent magnitude brighter than $m_{\rm F625W} < 26.3$.
The target fields of our sample galaxies comprise three galaxy clusters: MACSJ0717, MACSJ1149, and Abell 1689.
The two MACS clusters are part of the Hubble Frontier Field (HFF) clusters \citep{lotz17}.
%lensed galaxies with apparent magnitude cut of F625W $< 26.3$ and photometric redshift at $1.5 < z_{\rm phot} < 3.5$ in the three galaxy clusters: MACSJ0717, MACSJ1149, and Abell 1689.
From the Keck observations, 23 LAEs with the rest-frame equivalent width of Ly$\alpha$ (\ewlya ) $> 20 \ \angstrom$\footnote{This definition of a LAE with \ewlya\ $> 20 \ \angstrom$ is commonly adopted in the literature \citep[e.g.,][]{stei11,hath16}.} were found at spectroscopic redshifts $1.7 < z_{\rm spec} < 3.3$, with a median redshift of 2.43.

The median lensing magnification factor $\mu$ of the sample galaxies is 8.8, representing the ratio of the observed flux to the intrinsic flux without lensing.
This value is derived from strong lensing models for the clusters as reported by \cite{limo07}, \cite{limo16}, and \cite{jauz16} for Abell 1689, MACSJ0717, and MACSJ1149, respectively.
The R.A. and Dec. coordinates and the lensing magnification factor of the sample galaxies are listed in Table 1. 
We refer the readers to \cite{snap23} for further details about our sample LAEs.

%============Table 1===============
\begin{deluxetable*}{ccccccccccc}
\label{tab1}
%% Keep a portrait orientation

%% Over-ride the default font size
%% Use Default (12pt)

%% Use \tablewidth{?pt} to over-ride the default table width.
%% If you are unhappy with the default look at the end of the
%% *.log file to see what the default was set at before adjusting
%% this value.

%% This is the title of the table.
\tablecaption{The Physical Properties of the Sample LAEs.}

%% This command over-rides LaTeX's natural table count
%% and replaces it with this number.  LaTeX will increment 
%% all other tables after this table based on this number
\tablenum{1}

%% The \tablehead gives provides the column headers.  It
%% is currently set up so that the column labels are on the
%% top line and the units surrounded by ()s are in the 
%% bottom line.  You may add more header information by writing
%% another line between these lines. For each column that requries
%% extra information be sure to include a \colhead{text} command
%% and remember to end any extra lines with \\ and include the 
%% correct number of &s.

%---Old one%\tablehead{\colhead{Galaxy ID} & \colhead{Filter\tablenotemark{a}} & \colhead{Exposure time (s)} & \colhead{Rest-frame features\tablenotemark{b,c}}} 

\tablehead{\colhead{Galaxy ID} & \colhead{R.A. (deg)} & \colhead{Decl. (deg)} & \colhead{$z_{\rm spec}$} & \colhead{$M_{\rm UV}$\tablenotemark{a}} & \colhead{$r_{\rm eff}$ (pc)\tablenotemark{b}} & \colhead{$n_{\rm S}$\tablenotemark{c}} & \colhead{$b/a$\tablenotemark{d}} & \colhead{EW(Ly$\alpha$) ($\angstrom$)\tablenotemark{e}} & \colhead{$\mu$ \tablenotemark{f}}
}

%% All data must appear between the \startdata and \enddata commands
\startdata
A1689-257 & 197.8601135 & -1.358672653 & 1.705 & $-17.62^{+0.05}_{-0.05} $ & $116.8^{+5.07}_{-4.74}$ & $ 1.0 \pm 0.12 $ & $ 0.2 \pm 0.03 $ & $99.7^{+16.1}_{-16.1}$ & $6.19 \pm 0.25$ \\
A1689-280 & 197.8828396 & -1.357248833 & 1.705 & $-17.18^{+0.19}_{-0.21} $ &$ 178.67^{+25.52}_{-20.77} $ & $ 3.82 \pm 0.2 $ & $ 0.13 \pm 0.01 $ & $35.1^{+2.4}_{-2.4}$ & $22.08 \pm 3.55$ \\
A1689-539 & 197.8740075 & -1.352060819 & 3.046 & $-18.59^{+0.08}_{-0.08} $ &$ 141.53^{+6.45}_{-5.91} $ & $ 0.5 \pm 0.04 $ & $ 0.16 \pm 0.01 $ & $22.3^{+3.4}_{-3.4}$ & $8.47\pm 0.63$ \\
A1689-540 & 197.8709266 & -1.35206099 & 2.546 & $ -17.45^{+0.09}_{-0.09} $ &$ 77.47^{+6.0}_{-5.47} $ & $ 1.32 \pm 0.34 $ & $ 0.49 \pm 0.06 $ & $67.2^{+7.37}_{-8.74}$ & $9.04 \pm 0.61$ \\
A1689-830 & 197.8540603 & -1.344897081 & 2.666 & $-18.97^{+0.06}_{-0.06} $ &$ 93.45^{+4.97}_{-4.72} $ & $ 2.13 \pm 0.2 $ & $ 0.1 \pm 0.03 $ & $76.6^{+2.30}_{-2.27}$ & $5.6 \pm 0.27$ \\
A1689-920 & 197.8954312	& -1.34277496 & 2.546 & $-16.34^{+0.15}_{-0.17} $ &$ 23.27^{+3.49}_{-3.0} $ & $ 5.62 \pm 1.22 $ & $ 0.11 \pm 0.05 $ & $65.8^{+7.43}_{-6.28}$ & $25.12 \pm 3.0$ \\
A1689-946 & 197.8954312	& -1.34277496 &	2.287 & $-15.41^{+0.35}_{-0.39} $ &$ 42.27^{+13.44}_{-10.63} $ & $ 8.0 \pm 4.93 $ & $ 0.58 \pm 0.12 $ & $118.2^{+30.16}_{-24.46}$ & $30.2 \pm 6.0$ \\
A1689-1000 & 197.8777216 & -1.34024891 & 2.665 & $-13.69^{+11.22}_{-7.49} $ &$ 9.47^{+288.61}_{-9.41} $ & $ 1.03 \pm 0.05 $ & $ 0.11 \pm 0.02 $ & $66.7^{+2.82}_{-2.71}$ & $991 \pm 29824$ \\
A1689-1037 & 197.8966616 & -1.34028896 & 1.703 & $-15.79^{+0.35}_{-0.5} $ &$ 135.0^{+38.08}_{-22.44} $  & $ 2.09 \pm 0.12 $ & $ 0.49 \pm 0.02 $ & $23.4^{+3.69}_{-3.54}$ & $43.25 \pm 15.46$ \\
A1689-1117 & 197.895417	& -1.33847996 & 2.546 & $-10.03^{+12.9}_{-8.42} $ &$ 19.18^{+907.64}_{-19.13} $ & $ 0.5 \pm 0.22 $ & $ 0.13 \pm 0.03 $ & $216.5^{+62.94}_{-26.19}$ & $2334 \pm 141885$ \\
A1689-1197 & 197.8737117 & -1.335775444 & 1.706 & $-17.68^{+0.14}_{-0.16} $ &$ 186.89^{+16.97}_{-14.09} $ & $ 2.65 \pm 0.11 $ & $ 0.41 \pm 0.01 $ & $43.4^{+2.30}_{-2.30}$ & $17.22 \pm 2.15$ \\
A1689-40000 & 197.863348 & -1.347704975 & 1.837 & $-17.82^{+0.13}_{-0.14} $ &$ 174.09^{+17.03}_{-17.03} $ & $ 3.15 \pm 0.16 $ & $ 0.27 \pm 0.01 $ & $107.2^{+2.63}_{-2.61}$ & $25.82 \pm 2.1$ \\
A1689-40011 & 197.8651121 & -1.359898194 & 2.594 & $-17.54^{+0.16}_{-0.19} $ &$ 189.55^{+24.62}_{-20.55} $ & $ 2.42 \pm 0.2 $ & $ 0.34 \pm 0.01 $ & $160.4^{+8.78}_{-8.19}$ & $20.14 \pm 2.59$ \\
\hline
MACSJ1149-1098 & 177.4065351 & 22.39286171 & 1.895 & $-20.26^{+0.09}_{-0.1} $ &$ 203.06^{+1.9}_{-1.9} $ &  $ 5.17 \pm 0.21 $ & $ 0.05 \pm 0.0 $ & $52.1^{+3.49}_{-3.63}$ & $18.71^{+1.40}_{-1.78}$ \\
MACSJ1149-1721 & 177.4201963 & 22.40213713 & 2.095 & $-18.25^{+0.03}_{-0.04} $ &$ 276.83^{+14.09}_{-13.05} $ & $2.58\pm 0.36$ & $0.37\pm 0.03$
 & $197.5^{+33.45}_{-28.54}$ & $1.58^{+0.02}_{-0.02}$ \\
MACSJ1149-1856 & 177.3840617 & 22.40503821 & 2.035 & $-16.65^{+0.24}_{-0.24} $ & $ 203.06^{+62.98}_{-51.78} $ & $ 8.00 \pm 4.75 $ & $ 0.70 \pm 0.10 $ & $67.5^{+215.72}_{-96.03}$ & $2.83^{+0.17}_{-0.12}$ \\
MACSJ1149-2520 & 177.4068209 & 22.41625532 & 2.422 & $-19.59^{+0.07}_{-0.07} $ &$ 176.01^{+10.2}_{-9.8} $ & $8.00 \pm1.09$ & $0.85\pm 0.03$ & $110.1^{+18.73}_{-13.53}$ & $1.56^{+0.06}_{-0.04}$ \\
MACSJ1149-2533 & 177.4161338 & 22.41659674 & 3.227 & $-19.99^{+0.03}_{-0.03} $ &$ 590.16^{+12.27}_{-12.29} $ & $0.81 \pm 0.03$ & $0.45\pm0.01$ & $41^{+15.73}_{-14.26}$ & $1.32^{+0.02}_{-0.02}$ \\
MACSJ1149-2620 & 177.4120997 & 22.4187921 & 2.491 & $-17.92^{+0.04}_{-0.04} $ &$ 276.79^{+16.63}_{-16.16} $ & $1.36\pm 0.29$ & $0.21\pm 0.04$ & $25.9^{+34.36}_{-21.54}$ & $1.34^{+0.03}_{-0.03}$ \\
MACSJ1149-2761 & 177.4116563 & 22.42309318 & 3.31 & $-19.78^{+0.04}_{-0.04} $ &$ 479.89^{+15.27}_{-14.7} $ & $1.78\pm 0.12$ & $0.69\pm 0.02$ & $50^{+43.59}_{-32.32}$ & $1.33^{+0.03}_{-0.03}$ \\
\hline
MACSJ0717-744 & 109.3640224 & 37.73280612 & 3.042 & $-19.18^{+0.02}_{-0.02} $ &$ 268.13^{+6.37}_{-6.34} $ & $2.28\pm0.14$ & $0.1\pm0.01$ & $61.9^{+17.61}_{-15.62}$ & $1.70^{+0.02}_{-0.01}$ \\
MACSJ0717-1034 & 109.3653315 & 37.73842898 & 2.441 & $-18.67^{+0.03}_{-0.03} $ &$ 38.34^{+3.72}_{-3.7} $ & $7.3\pm1.44$ &$0.12\pm0.05$ & $63.2^{+13.37}_{-11.75}$ & $1.88^{+0.02}_{-0.01}$ \\
MACSJ0717-1170 & 109.3985407 & 37.74149598 & 1.859 & $-16.88^{+0.14}_{-0.16} $ &$ 83.0^{+7.74}_{-6.37} $ & $2.41\pm0.19$ &$0.5\pm 0.02$ & $73.6^{+5.52}_{-5.25}$ & $11.90^{+2.14}_{-0.95} $ \\
\hline
\hline
\enddata

\tablenotetext{a}{{The lensing magnification-corrected UV magnitude at rest-frame 1800 $\angstrom$ (Section \ref{sec:2_2}}).}
\tablenotetext{b}{{The lensing magnification-corrected, circularized  effective radius (Section \ref{subsec3_1:UV Sizes})}.}
\tablenotetext{c}{{The S\'ersic index (Section \ref{sec:2_2}}).}
\tablenotetext{d}{{The axis ratio (Section \ref{sec:2_2}).}}
\tablenotetext{e}{{The Ly$\alpha$ equivalent width (Section \ref{sec 2.1: LAE sample}).}}
\tablenotetext{f}{{The lensing magnification (Section \ref{sec 2.1: LAE sample}).}}

%% Include any \tablenotetext{key}{text}, \tablerefs{ref list},
%% or \tablecomments{text} between the \enddata and 
%% \end{deluxetable} commands

%% No \tablecomments indicated

%% No \tablerefs indicated
\end{deluxetable*}

%==================================

%;------2.2 Size Measurements------
\subsection{Size and Luminosity Measurements}
\label{sec:2_2}
We measure the UV size of our sample LAEs by fitting their 2D surface brightness profiles.
As described below, we utilize deep HST images that enable us to securely measure the UV size of our sample galaxies down to faint UV luminosity ($\simeq 0.01 L_{\rm *}$, where $L_{\rm *}$ is the characteristic UV luminosity corresponding to $M_{\rm UV} = -21$ at $z=3$, \citealt{redd09,stei18}). 
Thus, our analysis covers a wide range of UV luminosities for LAEs, including both bright 
($0.3-1$ $L_{\rm UV}/L_{*}$) and faint ($0.12-0.3$ $L_{\rm UV}/L_{*}$) luminosities.
Specifically, we use the HST Advanced Camera for Surveys/Wide Field Camera (ACS/WFC) F606W and F625W images for the sample galaxies in the two HFF clusters and the Abell cluster, respectively.
The images trace the rest frame UV-continuum ($1500 - 2200 \ \angstrom$) at redshifts $1.7 < z < 3.3$.
The F606W images are obtained from the HFF science products\footnote{https://archive.stsci.edu/prepds/frontier/}.
The images have a total exposure time of 27,015 and 24,816 seconds for MACSJ0717 and MACSJ1149 clusters, respectively, with a common pixel scale of 0.03$''$ $\rm pixel^{-1}$.
The F625W image for the Abell 1689 cluster was originally obtained from the HST-GO-9289 (PID: H. Ford) and has a total exposure time of 9500 seconds and a pixel scale of $0.04''$ $\rm pixel^{-1}$.
The F625W image we use in this paper was processed and presented in \cite{alav14,alav16}, and we refer the reader to these papers for further details on the image processing.

We employ the {\tt\string GALFIT} software \citep{peng02,peng10} to measure the size of our sample galaxies by fitting their 2D surface brightness profiles with a Sérsic light profile model \citep{sers68}.
We fit the image region centered on a galaxy, ensuring that the image size is sufficient to capture the galaxy's light while also allowing for an accurate estimation of the background sky light.
The point spread function (PSF) effect is taken into account in our surface brightness fitting by incorporating the PSF star image during the {\tt\string GALFIT} procedure.
We generated a PSF image for each cluster by selecting the unsaturated and isolated stars and stacking them using the {\tt\string StarFinder} \citep{diol00} procedure.

We put constraints on the fitting ranges of the structural parameters to prevent the fitting from resulting in unphysical results (e.g., unphysically large effective radius). 
Such constraints have been often adopted in other studies employing the least-$\chi^{2}$ fit algorithm for galaxy surface brightness such as {\tt\string GALFIT} \citep[e.g.,][]{peng02,vand12,kim16,shib19}.
Specifically, we consider the following fitting ranges: major-axis effective radius $r_{\rm maj}$ $>$ 0.5 pixels; S\'ersic index $0.5 < n_{\rm S} < 8$; axis-ratio $b/a > 0.1$, which are qualitatively similar to those of \cite{vand12}.

%We correct for lensing magnification to the size and luminosity measured in the image plane by using the lensing magnification factor derived from public strong lens models by \cite{limo07}, \cite{limo16}, and \cite{jauz16} for Abell 1689, MACSJ0717, and MACSJ1149 respectively.
We correct for lensing magnification to the size and luminosity measured in the image plane by using the magnification factor derived from the public strong lens models described in Section \ref{sec 2.1: LAE sample}.
For a summary of the lens models used here, we refer the reader to \cite{alav16}.

For the majority of sample galaxies (that is, 21 out of 23), their morphology is appreciably compact.
Thus, a simple lensing correction via dividing the size (that is, $r_{\rm maj}$ measured in the image plane) and luminosity by ($\sqrt{\mu}$) and $\mu$, respectively, can be regarded as a feasible way to estimate the overall intrinsic properties of a galaxy. 
This is to keep the surface brightness fixed, because lensing does not change it \citep[e.g.,][]{shar22}.

The remaining sample galaxies are two magnified arcs (Galaxy IDs: MACSJ1149-1098 and A1689-40000), for which we measure their sizes directly on the source-plane reconstructed images. This approach is necessary because their image-plane counterparts exhibit highly elongated arc morphologies, making the simple lensing corrections highly uncertain.

{We note that the reported UV magnitudes represent the magnitude at rest-frame 1800 $\angstrom$, which corresponds to the mean rest-frame wavelength of the imaging filters used at the redshifts of the sample galaxies. To derive this, we adopted} the UV spectral slope measured using the best-fit SED fitting technique described in {\cite{alav25}, and applied the slope to correct the observed magnitudes to rest-frame 1800 $\angstrom$}.
%We correct the galaxy magnitude for the mean wavelength of 1800 $\angstrom$ by adopting the UV spectral slope measured using the best SED fitting technique discussed in A. Alavi, 2025 (submitted to ApJ). 
The best SED fits are done similarly to what is presented in \citet{alav16}. In brief, the best SEDs for two HFF clusters, MACSJ0717 and MACSJ1149, utilize publicly available deep data in 10 {\it HST} broad-band filters, including F225W, F275W, F336W, F435W, F606W, F814W, F105W, F125W, F140W, and F160W. The UV data are sourced from {\it HST} program IDs 15940 (F225W; PI: Ribeiro) and 13389 (F275W and F336W; PI: B. Siana), while the optical and NIR data are obtained from {\it HST} program ID 13495 (PI: Lotz). For Abell 1689, we utilize publicly accessible photometric data in 8 {\it HST} bands, specifically  F225W, F275W, F336W, F475W, F625W, F775W, F814W, and F850LP. The UV data for A1689 are associated with program IDs 12201 and 12931 (PI: B. Siana). 

The physical properties of our sample galaxies are provided in Table 1.

%=====================Sec.3 Results========
%==========================================================
\section{Results} \label{sec3:retuls}

%-------------------------------------------------------------
%------Fig.1, histograms of size and luminosity, (and likely SigSFR)
\begin{figure*}
\includegraphics[width=\linewidth]{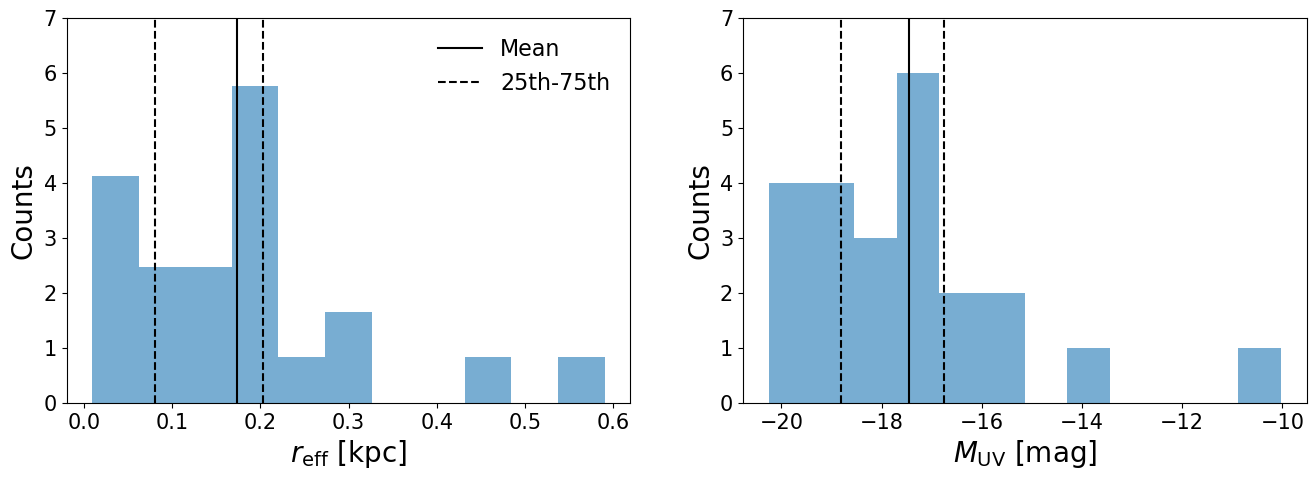}
\caption{\textbf{Left:} UV-continuum size (corrected for lensing) of our sample LAEs at $1.7 < z < 3.3$. The mean size (solid vertical line) and the interquartile range (dashed vertical lines) are $170^{+30}_{-90}$ pc.
\textbf{Right:} The lensing-corrected UV luminosity distribution for the LAEs. The median UV luminosity is $-17.66^{+0.91}_{-1.14}$ magnitudes, with the interquartile range represented in the same manner as in the left panel.
\label{fig1:Size distr}}
\end{figure*}

%-------------------------------------------------------------
%--------------------Section 3.1------------------------------
\subsection{Size Distribution of LAEs at Cosmic Noon $1.7 < z < 3.3$}
\label{subsec3_1:UV Sizes}
The main goal of this paper is to investigate the UV-continuum size of LAEs and compare the typical size of LAEs with that of continuum-selected star-forming galaxies (i.e., Lyman Break Galaxies and/or photometrically selected star-forming galaxies).
Our analysis extends previous size analyses of LAEs to fainter UV luminosity ($M_{\rm UV} \simeq -14$) by leveraging the gravitational lensing effects of the foreground clusters.

The left panel of Figure \ref{fig1:Size distr} shows the size distribution of our sample LAEs.
{Compared to the typical PSF FWHM of the HST images, the sizes of the majority of sample LAEs are resolved due to the high spatial resolution of the HST images. However, there are four galaxies (that is, A1689-540, A1689-920, A1689-946, and MACSJ1149-2520) whose morphology is so compact that the measured effective radius (in angular size) is typically about half the PSF FWHM ($\sim 2$ pixels).}
The size is corrected for lensing magnification, and measured as the circularized effective size \reff\ (that is, \reff = $r_{\rm maj} \times \sqrt{\rm b/a}$, where $b/a$
and $r_{\rm maj}$ are the galaxy axis ratio and effective radius along major-axis, respectively); This size definition is commonly adopted in other studies \citep[e.g.,][]{vand14,shib15,kim21,nedk24}, which allows us to compare our size measurements with the literature in a consistent manner.

The distribution shows the mean size of 170 pc and the associated standard deviation of 140 pc of the sample galaxies, with a long tail toward large sizes up to $\simeq 0.6$ kpc.
As we will further discuss below, our sample LAEs' size distribution indicates a very small typical size for their UV luminosity compared to typical (i.e., including all types of) star-forming galaxies with similar UV luminosity at similar redshifts.

Specifically, our UV bright ($0.3-1$ $L_{\rm UV}/L_{*}$) sub-sample galaxies show a factor of $\sim 4$ \textit{smaller} average size of $0.42 \pm 0.16$ kpc compared to those 
($1.4 \lesssim$ \reff\ $\lesssim 2$ kpc) of the photometrically selected star-forming galaxies (SFGs) and Lyman Break Galaxy (LBG) counterparts at similar redshifts $z=2-3$ \citep{bouw04,vand14,shib15,ribe16,nedk24}.
This trend of the small typical size of the sample LAEs is found regardless of specific UV luminosity bins; Consistent with the bright sub-sample, our faint ($0.12-0.3$ $L_{\rm UV}/L_{*}$) sub-sample galaxies also exhibits small sizes, with an average size of $0.30 \pm 0.18$ kpc. This contrasts with the average size of $\simeq 0.8$ kpc of the SFGs and LBGs counterparts at similar redshifts.\citep[e.g.,][]{vand14,shib15}.

{We note that while most studies referenced here report rest-frame UV sizes, some—namely \cite{vand14} for SFGs and LBGs, and \cite{redd22}, and \cite{liu23} for LAEs—measured rest-frame optical sizes (at $\sim 5000 \angstrom$). However, several studies suggest that the difference between UV and optical sizes is small: less than 10$\%$ for SFGs at $0.5<z<3$ \citep{nedk24}, and similarly small for LAEs at $z\sim 2-7$ \citep{shib19,song25}. In addition, the reported UV magnitudes span a rest-frame wavelength range from 1500 to 2800 $\angstrom$. Most studies use wavelengths between 1500 and 1900 $\angstrom$, with the exception of Shibuya et al. (2015), which reports UV magnitudes at 2800 $\angstrom$ for SFGs and LBGs. Assuming a typical UV slope of -1.7, representative of $z \sim 2$ LBGs (e.g., Hathi et al. 2009; Bouwens et al. 2009), we estimate that the magnitude difference between 1800 $\angstrom$ and 2800 $\angstrom$ is approximately 0.92 mag, and between 1800 $\angstrom$ and 1500 $\angstrom$ is about 0.44 mag across studies. Even with these modest differences in size measurements and UV magnitude wavelengths, the trend that LAEs are more compact than typical SFGs and LBGs at similar redshifts remains robust.}

While the sizes of our sample LAEs are smaller compared to typical SFGs at similar redshifts, they are consistent with the typical sizes of LAEs across a wide range of redshift $0.1 < z < 6$, where the reported size ranges between \reff $ = 0.3 - 1$ kpc \citep[e.g.,][]{dowh07,over08,bond09,tani09,bond12,malh12,jian13,paul18,shib19,rito19,kim21,flur22,redd22,liu23,ning24}.

The right panel of Figure \ref{fig1:Size distr} shows the UV luminosity distribution of our sample galaxies, ranging from bright (${M_{\rm UV} \simeq -20}$ to faint (${M_{\rm UV} \simeq -14}$) luminosities.
The median UV luminosity of these galaxies is -17.7.
The luminosities are corrected for lensing-magnification ($\mu$), as described in Section \ref{sec:sample and analysis}.

%=============Section 3.2 UV Size-Luminosity relation==========
\subsection{UV Size-luminosity Relations of LAEs}
\label{subsec3_2:UV Size-lum relation}
At all redshifts, galaxies show a tight relation between their size and luminosity (similarly stellar mass) \citep[e.g.,][]{bouw04,vand14,shib15,mori24,nedk24}, which is also supported by theoretical galaxy disk formation models \citep[e.g.,][]{fall80,barn87,mo98,wyit11,liu17}.
The size-luminosity relation is often parameterized as a power-law with a slope of $\alpha$:
\begin{eqnarray}
r_{\rm eff} = r_{0}\Bigg(\frac{L_{\rm UV}}{L_{0}} \Bigg)^{\alpha} ,
\label{Eq1}
\end{eqnarray}
where $L_{\rm UV}$, and $L_{0}$ are the UV luminosity of galaxies, a fiducial UV luminosity we take as the $z=3$ characteristic luminosity, (i.e., $L_0 = L_{\rm *, z=3}$, corresponding to $M_{\rm UV} = -21$), respectively.
Also, $r_{0}$ is defined to be the size at $L_{0}$.

%------Fig.2 UV Size-luminosity relation
\begin{figure}
\includegraphics[width=1\linewidth]{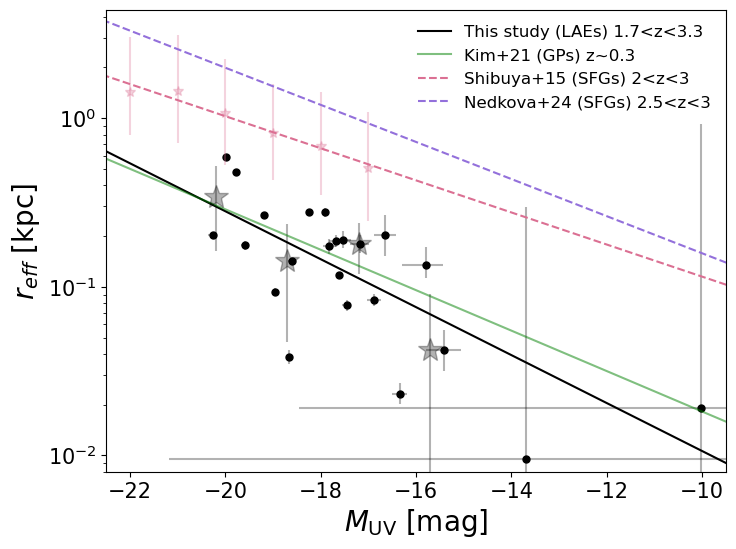}
\caption{
The UV size-luminosity relation for our sample of LAEs (black points) at $1.7 < z < 3.3$ (Section \ref{subsec3_2:UV Size-lum relation}). The black solid line represents the fitted relation for these galaxies, while the grey star points indicate the median sizes of our sample galaxies binned by UV luminosity.
Light pink points and dashed line indicate the same relation for continuum-selected star-forming galaxies (SFGs) at similar redshifts \citep{shib15}.
{Also, the light purple relation is taken from \cite{nedk24}, where their stellar mass-size relation was converted to the corresponding UV luminosity-size relation using the stellar mass-UV luminosity relation presented in \cite{shib15} for SFGs at $2 < z <3$.} 
The green solid line represents the relation for local Green Pea galaxies ($z \sim 0.3$), which are low-redshift analogs to high-redshift LAEs \citep{kim21}. 
At a given UV luminosity, our sample LAEs are smaller in size compared to typical SFGs, but they exhibit a size-luminosity relation similar to that of Green Pea galaxies, suggesting that LAEs maintain a consistently small size across redshifts.
\label{fig2:Size Lum relation}}
\end{figure}

We derive the size-luminosity relation of our sample LAEs in Figure \ref{fig2:Size Lum relation} by extending the previous analysis \citep[e.g.,][]{jian13,kim21} to fainter galaxies $M_{\rm UV} \simeq -14$. 
We exclude the two highly magnified galaxies (A1689-1000 and A1689-1117) from the fit due to their large uncertainties in the magnification factor (i.e., $\sigma( \mu)/ \mu > 1$, see Table 1).
The fitted slope and intercept are $\alpha = 0.36 \ \pm 0.12 $ and $r_{\rm 0} = 0.39^{+0.17}_{-0.12}$ kpc, respectively.
The slope of our sample galaxies is broadly consistent, within uncertainties, with that of typical SFGs and LBGs at $0.5 < z < 6$, which show a slope range of $0.15 \lesssim \alpha \lesssim 0.5$, with a typical value of $\alpha \simeq 0.27$ \citep{graz12,jian13,huan13,vand14,shib15,curt16,kawa18,nedk24}.

In contrast to the slope $\alpha$, the intercept ($r_{0}=390$ pc) of the size-luminosity relation for our sample LAEs is smaller than that observed for typical SFGs and LBGs ($r_{0} \simeq 1$ kpc).
The small size of LAEs (at a fixed UV luminosity) leads to a high star formation surface density (i.e., star formation rate per unit area, $\Sigma$SFR).
We will further discuss how the high $\Sigma$SFR of LAEs is related to the escape of \lya\ photons in Section \ref{sec4:Discussion}.

%==============Section 3.3 Relations between UV size and LyA emisson
\subsection{The relation between UV size and the \lya\ Equivalent Width} 
\label{subsec3_3:UV Size EW(LyA) Relations}
\begin{figure}
\includegraphics[width=\linewidth]{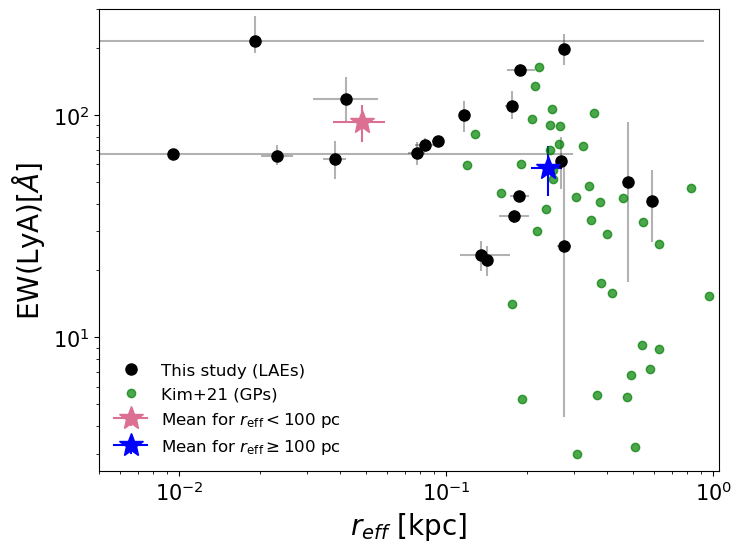}
\caption{The figure illustrates the relation between \lya\ emission equivalent width (\ewlya ) and UV-continuum size (\reff ) for LAEs, with LAEs shown as black dots.
The thick pink point represents the mean \reff\ and \ewlya\ for the small-size sub-sample (\reff\ $< 100$ pc), while the blue point corresponds to the large-size sub-sample (\reff\ $\geq 100$ pc). 
The green points represent the relation for Green Pea galaxies \citep{kim21} for comparison.
The data distribution and median values indicate that smaller sizes are generally associated with higher \ewlya, which is consistent with previous studies (Section \ref{subsec3_3:UV Size EW(LyA) Relations}).
\label{fig3:Size EWLyA}}
\end{figure}

In this section, we investigate the possible correlations between UV size (\reff ) and the \lya\ equivalent width (\ewlya ) in our sample LAEs.
We also compare the trends of our sample galaxies with those seen in other LAEs from the literature.
Figure \ref{fig3:Size EWLyA} shows the correlations between \reff\ and \ewlya.
There appears to be an anti-correlation between \reff\ and \ewlya, with moderate scatter, where smaller galaxies tend to have \textit{higher} \ewlya.
While the small sample size may impact the reliability of the Spearman correlation test, leading to a weak correlation ($p=0.09$, with $p \leq 0.05$ considered significant), the mean \ewlya\ of the small size (\reff\ $< 0.1$ kpc) sub-sample galaxies is higher than that of the large size (\reff\ $\geq 0.1$ kpc) counterparts, which values of $93 \pm 17$ $\angstrom$ and $58 \pm 15$ $\angstrom$, respectively.

%------Fig.3 EW(LyA) vs. Size 

A qualitatively similar anti-correlation between UV size and \ewlya\ has been found in other LAEs across a wide range of redshift (0.1 $\lesssim z \lesssim 7$) \citep[e.g.,][]{bond09,bond12,guai15,paul18,kim21,puch22,redd22,keru22}.
Specifically, at redshifts $z \simeq 2.1$ and 3.1 similar to our sample LAEs, 
\cite{bond12} showed that high \ewlya\ LAEs have a smaller median UV size compared to low \ewlya\ LAEs.
Additionally, a linear-fit of the UV size and \ewlya\ relation for a sample of high-\textit{z} LAEs at $2 \lesssim z \lesssim 6$, as reported by \cite{paul18}, reveals a negative slope of $(-3.5 \pm 1.2) \times 10^{-3}$. This result seems qualitatively consistent with the anti-correlation observed in our sample LAEs, where the linear fit between \reff\ and \ewlya\ shows a negative slope of $(-5.0 \pm 5.0) \times 10^{-4}$.

At lower redshifts  ($0.03 < z < 0.2$), the Lyman alpha reference sample (LARS) reported a significant anti-correlation with the Spearman correlation coefficient ($p$-value) of $-0.64$ (0.03) based on 12 local star-forming galaxies \citep{guai15}. 
Similarly, \cite{kim21} reported a significant anti-correlation ($p$-value $< 0.001$) between \ewlya\ and UV size based on a sample of Green Pea galaxies.
This suggests that a small size is preferred for significant \lya\ emission.

%=============Section 4 Discussion
\section{Discussion} \label{sec4:Discussion}
\subsection{Small Size and High $\Sigma$SFR as a Crucial Condition for \lya\ Escape}
\label{subsec4_1:high_Sigma_SFR}
Our UV size analysis of LAEs at Cosmic Noon reveals that {their sizes are approximately 3-4$\times$ smaller} than those of continuum-selected SFGs and that they follow a distinct size-luminosity relation compared to typical SFGs.
This is consistent with other studies analyzing the morphology of LAEs across a wide range of redshift ($0.1 \lesssim z \lesssim 7$) \citep{malh12,jian13,izot16,izot18a,kim20,kim21,flur22,liu23,ning24}.
Our analysis extends the LAE morphology analysis to faint UV luminosity ($M_{\rm UV} \simeq -14$).

\begin{figure}
\includegraphics[width=\linewidth]{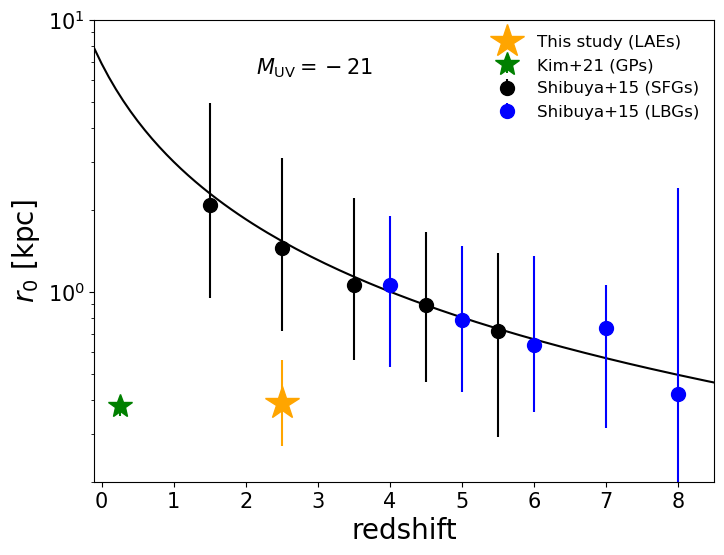}
\caption{The intercept ($r_{\rm 0}$, size at $M_{\rm UV} = -21$) of the size-luminosity relation for our sample LAEs (indicated by the orange star) is distinctly smaller compared to that of typical (continuum-selected) star-forming galaxies at similar redshifts ($\sim 0.4 $ vs. $\sim 1.5$ kpc), and does not seem to follow the expected redshift-dependent size growth seen in SFGs. 
For comparison, the $r_{\rm 0}$ of Green Pea galaxies is shown as a green star, which closely matches the $r_{\rm 0}$ of our sample LAEs at Cosmic Noon.
\label{fig4:Size r0 no evolution}}
\end{figure}

Due to their small sizes, LAEs show a small intercept (i.e., $r_{\rm 0} \simeq 0.4$ kpc, the size at $M_{\rm UV}=-21$ as in Eq. \ref{Eq1}) in the size-luminosity relation, resulting in LAEs lying below the relation compared to their SFG counterparts (Section \ref{subsec3_2:UV Size-lum relation}).
Interestingly, the small $r_{\rm 0}$ of our sample LAEs does not seem to follow the redshift evolution of $r_{\rm 0}$ for continuum-selected SFGs and LBGs, but is instead consistent with that of low-redshift ($z \sim 0.3$) LAEs (also known as Green Pea galaxies) \citep{izot16,izot18a,kim21,flur22}.
This is shown in Figure \ref{fig4:Size r0 no evolution}.
The distinctly small intercept size $r_{\rm 0}$ for the size-luminosity relations of LAEs both at low and high-redshift ($z \sim 0.3-3.5$) corroborates that LAEs have characteristic compact sizes independent of redshift \citep{malh12,jian13} and that the compact UV size likely plays a key role in the escape of \lya\ emission (Figure \ref{fig3:Size EWLyA}) \citep{kim20,kim21}.

%------Fig.4 UV-luminosity relation, r0 of our sample LAEs with the literature, redshift evolution

%------Fig.5 \Sigma SFR vs. Redshift with the literature
\begin{figure*}
\includegraphics[width=\linewidth]{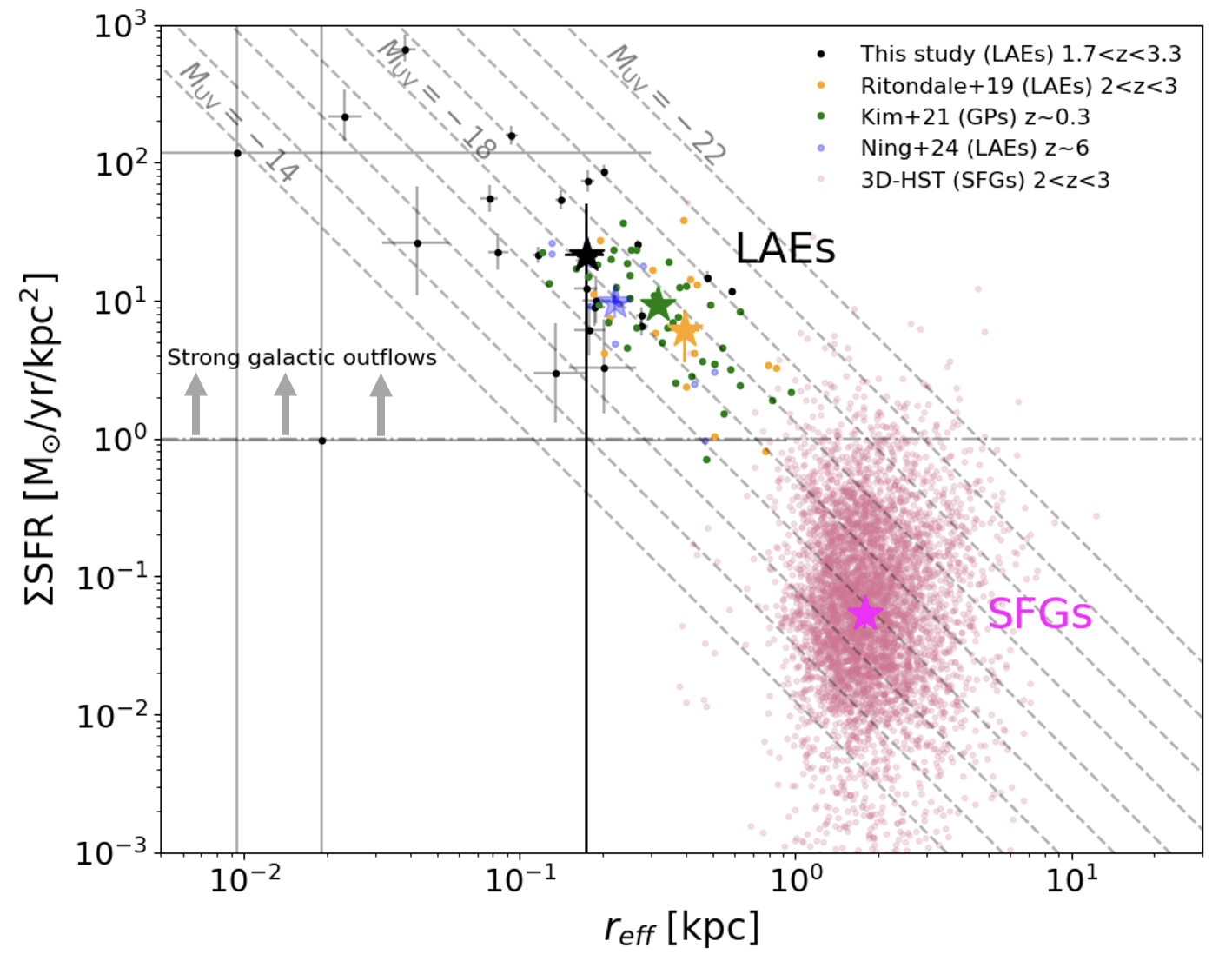}
\caption{
The Star Formation Surface Density ($\Sigma$SFR) vs. Size (\reff ) relation for LAEs compared to typical star-forming galaxies (SFGs) is shown. 
The black circles represent our sample of LAEs, with the median denoted by the black star. LAEs from the literature, spanning a redshift range of $0.3 \lesssim z \lesssim 6$ \citep{rito19,kim21,ning24}), are plotted as green, blue, and orange points. 
For comparison, typical SFGs from the 3D-HST survey are shown as pink points \citep{skel14,whit14}. 
The dashed lines correspond to constant UV luminosity, ranging from $M_{\rm UV} = -22$ to $M_{\rm UV} = -14$. 
{The dot-dashed horizontal line marks an empirical $\Sigma$SFR threshold of $= 1$ $M_{\sun} \ \rm{yr}^{-1} \ \rm{kpc^{-2}}$, derived from the fact that the majority of LAEs—87 out of 91 (96$\%$)—compiled from this study and the literature lie above this value. 
This empirical threshold is an order of magnitude higher than the $\Sigma$SFR threshold of $= 0.1$ $M_{\sun} \ \rm{yr}^{-1} \ \rm{kpc^{-2}}$ suggested by \cite{heck01,shar17} as indicative of conditions for strong galactic outflows (Section \ref{subsec4_1:high_Sigma_SFR}).}
Given their small size for a given UV luminosity, LAEs are characterized by compact morphologies and exhibit significantly higher $\Sigma$SFR) compared to typical SFGs—by two orders of magnitude. 
This suggests that compact morphology plays a crucial role in facilitating the escape of \lya\ photons. 
Additionally, the distinctive location of LAEs in the $\Sigma$SFR vs. size diagram can be used to identify potential \lya -emitters.
\label{fig5:Sigma SFR of LAE vs. SFGs}}
\end{figure*}

Notably, the very small size of LAEs for a given UV luminosity results in concentrated star formation activity per unit area{---that is, high star formation surface density ($\Sigma$SFR), defined as: $\Sigma$SFR $\equiv \frac{\rm{SFR}}{2 \pi r_{\rm{eff}}^{2}}  \left (\frac{M_{\sun} \ \rm{yr}^{-1}}{{\rm kpc}^2} \right )$. Such high $\Sigma$SFR}
is often associated with the presence of compact star-forming clumps within the galaxies \citep[e.g.,][]{keel05,kim20,vanz22,kim23,nava24,owen24}.
Indeed, due to their high $\Sigma$SFR, LAEs are clearly separated from the typical SFGs in the diagram of $\Sigma$SFR vs. size, as shown in Figure \ref{fig5:Sigma SFR of LAE vs. SFGs}. 

In the figure, we derive the SFR of our sample galaxies using the SFR-UV continuum flux relation from \cite{kenn98}, with the SFG counterparts taken from the 3D-HST survey \citep{skel14,whit14}. 
{For the SFG counterparts, the SFR is measured from the SED fitting using the FAST package \citep{krie09}. We also note that their \reff\ is measured in the rest-frame optical\footnote{{Although their size measurements are in the rest-frame optical—unlike this study, which uses UV measurements—we note that the difference between UV and optical sizes is typically small ($< 10 \%$; see Figure 7 in \cite{nedk24}). Therefore, using optical \reff\ values for the SFG counterparts does not affect the overall trend showing that LAEs are more compact than typical SFGs, as seen in Figure \ref{fig5:Sigma SFR of LAE vs. SFGs}.}}}.
Specifically, our sample LAEs with a mean \reff\ of $170 \pm 140$ pc show $\Sigma$SFR $\gtrsim 1$ $M_{\sun} \ \rm{yr}^{-1} \ \rm{kpc^{-2}}$, which is more than an order of magnitude higher than that of typical SFGs with similar UV luminosities and redshifts, as is clear from the lines of constant UV-magnitudes \citep[e.g.,][]{skel14,whit14}.
Consistent with our sample of LAEs, other LAEs from the literature \citep[i.e.,][]{rito19,kim21,ning24}\footnote{{The SFR reported in \cite{rito19} and \cite{kim21} are based on the same relation between SFR and UV continuum flux. However, \cite{rito19} adopted a different conversion coefficient of $1.12 \times 10^{-28}$ (from \cite{mada14}), compared the coefficient of $1.4 \times 10^{-28}$ used by \cite{kim21} and this study.
Additionally, the SFR in \cite{ning24} is derived from the SED fitting using the {\tt\string Bagpipes} code.}} also exhibit small sizes and high $\Sigma$SFR, occupying a distinct position in the $\Sigma$SFR vs. \reff\ parameter space compared to SFGs.

{Why do LAEs typically show small size (\reff\ $\sim 0.3$ kpc, which is the median \reff\ of all the LAEs, including those in this study and from other studies shown in Figure \ref{fig5:Sigma SFR of LAE vs. SFGs})} and high $\Sigma$SFR ($\gtrsim 1$ $M_{\sun} \ \rm{yr}^{-1} \ \rm{kpc^{-2}}$) compared to the SFG counterparts with similar UV luminosity, regardless of redshifts?
Is the compact morphology of LAEs related to the escape of \lya\ photons?
It has been suggested that a high $\Sigma$SFR indicates the presence of strong galactic outflows/winds. In this context, stellar winds and supernovae (SN) feedback in a dense starburst region generate significant radiation pressure, resulting in a galactic-scale outflow. \citep[e.g.,][]{meur97,heck01,heck01b,heck15,alex15,heck16,shar17,cen20,meno24}.
Indeed, $\Sigma$SFR shows a positive correlation with gas pressure in HII regions \citep{meur97,kim11,jian19b}, high ionization states (i.e., O32 line ratio), and electron density \citep{redd23a,redd23b}, suggesting that high $\Sigma$SFR is closely connected to the extreme ISM conditions resulting from galactic outflows.

{Specifically, a positive correlation between $\Sigma$SFR and thermal gas pressure in HII regions is expected if thermal and radiation pressure are primarily driven by stellar feedback. For instance, mechanical energy injection from stellar winds and/or supernovae in dense star-forming regions can increase the thermal pressure \citep{stri09}.}
{\cite{heck90} also show that, in starburst galaxies with strong galactic outflows, the pressure is often dominated by thermal pressure. The thermal pressure of ionized gas can be approximated as $P = N_{\rm total}Tk_{\rm B} \simeq 2n_{\rm e}Tk_{\rm B}$, where $n_{\rm e}$, $T$, $k_{\rm B}$ are the electron density and the electron temperature, and the Boltzmann constant, respectively. Therefore, a positive correlation between $\Sigma$SFR and $n_{\rm e}$ is also expected, particularly given that $T$ typically remains within a relatively narrow range for SFGs (10,000$-$20,000 K; e.g., \citealt{welc24,sand25}).}

{The observed correlation between $\Sigma$SFR and ionization state—as quantified by the O32 line ratio, defined as ${\rm O32} \equiv \frac{\OIII\ 5007}{\OII\ 3727,3729}$  —may also be explained by the presence of compact, young, and massive star clusters. These clusters produce intense ionizing UV radiation, photoionizing the surrounding ISM and resulting in a high ionization parameter ($U$). Indeed, in the local universe, super star clusters have been observed to exhibit elevated ionization states, with typical values of ${\rm log}U \simeq -2.3$ \citep[e.g.,][]{inde09,leit18,mich19}.
Furthermore, at Cosmic Noon, spatially resolved analyses of the strongly lensed Ly$\alpha$-emitting and LyC-leaking galaxy at $z=2.37$—also known as the Sunburst Arc—have revealed a remarkable spatial coincidence of high $\Sigma$SFR, high ionization state, and evidence of outflows \citep{main22,kim23,pasc23}. 
Specifically, the compact LyC-leaking star-forming region--likely hosting a super star cluster--in this system shows a high O32 ($\sim 11$) and clear signs of outflows, as indicated by a blueshifted, broad emission component (FWHM = $327 {\rm km/s}$) in the \OIII $\lambda 5007$ line. In contrast, the non-leaking star-forming regions within the galaxy lack these signatures.
}

Considering $\Sigma$SFR threshold of $> 0.1$ $M_{\sun} \ \rm{yr}^{-1} \ \rm{kpc^{-2}}$ for having galactic outflows, as suggested by \cite{heck01,shar17}, it is notable that all of our sample LAEs exceed this threshold, as shown in Figure \ref{fig5:Sigma SFR of LAE vs. SFGs}. 
Furthermore, most of our sample LAEs, along with others from the literature, have $\Sigma$SFR values well above this threshold, typically exceeding $\Sigma$SFR $\gtrsim 1$ $M_{\sun} \ \rm{yr}^{-1} \ \rm{kpc^{-2}}$. 
{This suggests that strong galactic outflows could be present in LAEs across a range of redshifts.}
{For instance, based on a sample of local starburst galaxies and Lyman Break Analogs (LBAs),  \cite{alex15} (see also \cite{heck15}) demonstrated a strong correlation between $\Sigma$SFR and outflow velocity, as measured from the UV Si III absorption line, which traces the ionized gas. They reported a Pearson correlation coefficient of 0.66 and an associated $p$-value of 0.001.}

Strong galactic outflows have been considered as one of the promising mechanisms to create under-dense channels in the ISM for the escape of \lya\ and potentially LyC photons \citep{keel05,bort14,alex15,main22,amor24}.
Indeed, Green Pea galaxies--which are low-redshift ($z \sim 0.3$) LAEs--are characterized with compact morphology (high $\Sigma$SFR) \citep{kim21}, and show high gas pressure \citep{jian19b} and outflows \citep{amor24}.
Notably, \cite{kim20} demonstrated direct correlations between central $\Sigma$SFR (and specific $\Sigma$SFR, $\Sigma$sSFR = $\Sigma$SFR/$M_{\rm star}$) and the \lya\ emission line properties.
They found significant correlations between $\Sigma$sSFR and \ewlya\ and \lya\ escape fraction, suggesting that an intense central starburst can drive galactic outflows in galaxies with shallow gravitational potential wells, thus clearing channels for the escape of \lya\ photons.
Qualitatively consistent conclusions on the importance of $\Sigma$SFR and specific $\Sigma$SFR on the escape of \lya\ and LyC have been reported based on studies of Green Pea galaxies and LAEs at $1.8 \lesssim z \lesssim 3.5$ \citep[e.g.,][]{redd22,flur22,puch22,jask24}.
{However, it should be noted that high $\Sigma$SFR alone may not be a sufficient condition for the presence of strong outflows. For instance, \cite{carr25} find that compact galaxies seem to show faster outflows in ionized gas but may lack outflows in neutral or low-ionization gas. Also, some \lya -emitting GPs do not show fast outflows \citep{jask17}. Both studies suggest that the age of the starburst may affect whether or not outflows are present in compact galaxies.}

The presence of strong outflows and the clearing of channels in the ISM for \lya\ escape are further supported by the weak low-ionization interstellar (LIS) absorption lines observed in our sample LAEs \citep{snap24} as well as other LAEs \citep[e.g.,][]{shap03,stei11,henr15}, as LIS absorption lines serve as an effective tracer of the neutral hydrogen covering fraction.
{However, we also note that there are other possible explanations for the low gas covering fraction that facilitates the escape of \Lya\ photons. For instance, condensation and fragmentation of cool gas clouds—driven by catastrophic cooling during the early stages of a starburst—could lead to a reduced covering fraction \citep{carr25}. \cite{carr25} also argue that supernova-driven outflows could actually increase gas covering fraction by lifting cool clouds. Additionally, \cite{haye23} suggest that low covering fraction may be attributed by photoionization, rather than outflows.}

Our results on the compact UV size and high $\Sigma$SFR of a sample of 23 LAEs at Cosmic Noon emphasize the importance of compact morphology on the escape of \lya\ photons, in agreement with galactic outflow-driven \lya\ escape mechanisms. {However, it should be noted that other factors may also contribute to low gas covering fractions, such as the condensation and fragmentation of cool gas clouds, or photoionization effects, as discussed above.}

%=============Section 5. Summary and Conclusionss
\section{Summary And Conclusions} \label{sec5:summary and conclusions}
We investigate the UV-continuum size and luminosity of a sample of 23 \lya -emitters at $1.7 < z < 3.3$. Using deep, high-resolution HST imaging combined with gravitational lensing from a foreground galaxy cluster, we can robustly measure the morphology of our sample galaxies down to faint UV luminosities ($M_{\rm UV} \simeq -14$) at Cosmic Noon. 
We compare the sizes of these LAEs with those of continuum-selected star-forming galaxies (SFGs) at similar redshifts. 
Our analysis reveals that the average size of LAEs is distinctly smaller than that of typical SFGs, reinforcing previous findings of the characteristic small size of LAEs, which appears to be independent of redshift.

Our primary conclusions are summarized below.

\begin{itemize}
\item 
Our sample of LAEs exhibits very small sizes, with a mean effective radius (\reff ) of 170 pc. The size distribution (Figure \ref{fig1:Size distr}) is narrow, with a standard deviation of 140 pc, though there is a long tail toward larger sizes (up to $\simeq 600$ pc). 
Compared to continuum-selected normal SFGs at similar redshifts, the typical sizes of our LAEs are approximately a factor of $\sim 3$ smaller at a given UV luminosity (Section 3.1). 
The small sizes observed in our sample are consistent with those of LAEs at similar redshifts, as well as with low-redshift Green Pea galaxies. This suggests that the compact morphology and small sizes of LAEs play a key role in facilitating the escape of \lya\ photons.

\item The UV size-luminosity relation of LAEs (Figure \ref{fig2:Size Lum relation}) shows a fitted slope $\textbf{$\alpha$}$ of $0.36 \pm 0.12$ and an intercept of $r_{\rm 0} = 0.39^{+0.17}_{-0.12}$ kpc measured at $M_{\rm UV} = -21$. 
{The slope measured for our sample of LAEs is slightly steeper than the average value ($\alpha \sim 0.27$) typically observed in star-forming galaxies at both low and high redshifts. However, it is consistent with the typical range of slope values for star-forming galaxies (i.e., $0.15 \lesssim \alpha \lesssim 0.5$), within the uncertainties (Section \ref{subsec3_2:UV Size-lum relation}).}
%\textbf{The slope of our sample LAEs is slightly higher than the average slope values $\alpha \sim 2.7$ of typical star-forming galaxies at low and high redshifts. However, it is consistent with the ranges of slope values of typical SFGs (i.e., $0.15 \lesssim \alpha \lesssim 0.5$), within the uncertainties (Section \ref{subsec3_2:UV Size-lum relation}).}

\item Unlike typical SFGs, the intercept $r_{\rm 0}$ of LAEs does not show the expected size growth with decreasing redshift (Figure \ref{fig4:Size r0 no evolution}). 
Specifically, LAEs have an $r_{\rm 0}$ that is approximately {$3-4 \times$ smaller} than that of SFGs at similar redshifts. This smaller $r_{\rm 0}$ is consistent with the sizes observed in low-$z$ Green Pea galaxies (i.e., low-$z$ LAEs. {However, it should be noted that Green Pea galaxies likely represent only a subset of LAEs with particularly strong optical emission lines and may not fully represent the broader low-$z$ LAE population due to their initial selection based on strong optical lines (See Section \ref{sec:intro})}). This suggests that LAEs maintain compact, non-evolving sizes regardless of redshift.

\item There is an anti-correlation between UV-continuum size (\reff ) and \ewlya\ (Figure \ref{fig3:Size EWLyA}), where \ewlya\ decreases as \reff\ increases. 
This trend is consistent with previous studies of LAEs across a wide redshift range ($0.1 < z < 6$), suggesting that a compact size is favorable for significant \lya\ emission.

\item 
Due to their compact size for a given UV luminosity, our sample of LAEs exhibits high $\Sigma$SFR (1$-$600 $M_{\sun} \ \rm{yr}^{-1} \ \rm{kpc^{-2}}$), which is more than two orders of magnitude higher than that of continuum-selected SFGs at similar redshifts. 
With a $\Sigma$SFR threshold of $> 0.1$ $M_{\sun} \ \rm{yr}^{-1} \ \rm{kpc^{-2}}$ for galactic outflows, {our sample LAEs surpass this threshold, indicating favorable physical conditions for strong outflows in these galaxies.}
Moreover, other LAEs in the literature show similar small sizes and high $\Sigma$SFR values, reinforcing the trend.

We find that LAEs occupy a distinct region in the $\Sigma$SFR vs. \reff \ diagram, with most LAEs having smaller sizes and higher $\Sigma$SFR for a given UV luminosity compared to SFG counterparts (Figure \ref{fig5:Sigma SFR of LAE vs. SFGs}). This suggests that the $\Sigma$SFR vs. size relationship can be a useful tool for identifying \lya -emitters.

\end{itemize}
In conclusion, our results suggest that compact morphology and high $\Sigma$SFR are key factors in enabling \lya\ (and potentially LyC) photon escape, as they create favorable conditions for strong galactic outflows. These outflows, likely driven by stellar feedback and/or supernovae from concentrated starburst regions, create under-dense channels that facilitate \lya\ escape. 

Our study implies that small sizes for a given luminosity can be an effective criterion for selecting \lya -emitters.

%==============Figure explanation============================
%\subsection{Figures\label{subsec:figures}}
%% The "ht!" tells LaTeX to put the figure "here" first, at the "top" next
%% and to override the normal way of calculating a float position
%\begin{figure}[ht!]
%\plotone{samplefig.png}
%\caption{The cost for an author to publish an article has trended downward
%over time. This figure shows the average cost of an article from 1990 to 2020 in 2021 adjusted dollars. 
%\label{fig:general}}
%\end{figure}
%=========================================================

%% Also note that the akcnowlodgment environment does not support long amounts of text. If you have a lot of people and institutions to acknowledge, do not use this command. Instead, create a new \section{Acknowledgments}.
\begin{acknowledgments}
We thank the referee for constructive comments that improved the quality of the manuscript.
KJK thanks Yu-Heng Lin, Sangeeta Malhotra, and James E. Rhoads for their useful discussions on this study. 
Support for this work was provided by awards HST-GO-15940 from STScI, which is operated by AURA, Inc. for the National Aeronautics Space Administration (NASA) under contract NAS 5-26555.
\end{acknowledgments}


\vspace{5mm}
\facilities{HST(ACS/WFC)}


\software{astropy \citep{Astropy2013,Astropy2018,Astropy2022}
          }



%%%%\appendix

%%%\section{Appendix information}
%Appendices can be broken into separate sections just like in the main text.
%The only difference is that each appendix section is indexed by a letter
%(A, B, C, etc.) instead of a number.  Likewise numbered equations have
%the section letter appended.  Here is an equation as an example.
%\begin{equation}
%I = \frac{1}{1 + d_{1}^{P (1 + d_{2} )}}
%\end{equation}
%Appendix tables and figures should not be numbered like equations. Instead
%they should continue the sequence from the main article body.


%\bibliography{sample631}{}
%\bibliographystyle{aasjournal}

\citestyle{aa}
\bibliographystyle{aasjournal}



\end{document}